\documentclass{aastex}          
\usepackage{spr-astr-addons}    
\usepackage{graphicx}
\usepackage{wrapfig}
\usepackage{amsmath}
\usepackage{amssymb}
\usepackage{bm}
\usepackage{latexsym}
\usepackage{color}

\begin{document}
%
\title{Kelvin--Helmholtz instability of magnetohydrodynamic waves propagating on solar surges}

\shorttitle{KH instability in solar surges}
\shortauthors{Zhelyazkov et al.}

\author{I.~Zhelyazkov} \affil{Faculty of Physics, Sofia University, 1164 Sofia, Bulgaria}
\and
\author{R.~Chandra}
\affil{Department of Physics, DSB Campus, Kumaun University, Nainital 263\,002, India}
\and
\author{A.~K.~Srivastava}
\affil{Department of Physics, Indian Institute of Technology (Banaras Hindu University), Varanasi 221\,005, India}
\and
\author{T.~Mishonov} \affil{Faculty of Physics, Sofia University, 1164 Sofia, Bulgaria}


\begin{abstract}
In the present paper, we study the evolutionary conditions for Kelvin--Helmholtz (KH) instability in a high-temperature solar surge observed in NOAA AR11271 using the \emph{Solar Dynamics Observatory\/} data on 2011 August 25. The jet with speed of ${\approx}100$~km\,s$^{-1}$, width of $7$~Mm, and electron number density of $4.17\times 10^{9}$~cm$^{-3}$ is assumed to be confined in an untwisted/twisted magnetic flux tube with magnetic field of $10$~G.  The temperature of the plasma flow is $2 \times 10^6$~K while that of its environment, according to the observational data, is of the order of $10^6$~K.  The electron number density of surrounding magnetized plasma is evaluated to be equal to $1.15 \times 10^{9}$~cm$^{-3}$.  Under these conditions, the Alfv\'en speed inside the flux tube is $337.6$~km\,s$^{-1}$, the sound speed is around $166$~km\,s$^{-1}$, while these characteristic speeds of the environment are ${\cong}719$~km\,s$^{-1}$ and ${\cong}117$~km\,s$^{-1}$, respectively.  We study the propagation of normal MHD modes in the flux tube considering the two cases, notably of untwisted magnetic flux tube and the twisted one.  The numerical solution to the dispersion relation shows that the kink ($m = 1$) wave traveling in an untwisted flux tube becomes unstable if the jet speed exceeds $1060$~km\,s$^{-1}$ -- a speed which is inaccessible for solar surges.  A weak twist (the ratio of azimuthal to longitudinal magnetic field component) of the internal magnetic field in the range of $0.025$--$0.2$ does not change substantially the critical flow velocity.  Thus, one implies that, in general, the kink mode is stable against the KH instability.  It turns out, however, that the $m = -2$ and $m = -3$ MHD modes can become unstable when the twist parameter has values between $0.2$ and $0.4$. Therefore, the corresponding critical jet speed for instability onset lies in the range of $93.5$--$99.3$~km\,s$^{-1}$.  The instability wave growth rate, depending on the value of the wavelength, is of the order of several dozen inverse milliseconds.  It remains to be seen whether these predictions will be observationally validated in future in the coronal jet-like structures in abundant measure.
\end{abstract}


%
\section{Introduction}
\label{sec:intro}

Solar surges seen in H$\alpha$ are cool plasma ejections from chromospheric to coronal heights.  Preceded by H$\alpha$ brightening, chromospheric mass is ejected along nearly straight trajectories with the typical velocities of $10$--$200$~km\,s$^{-1}$ from the H$\alpha$ brightening (roots of surges).  The ejected material moves upward to the heights up to $200$~Mm, and then either fades or returns back to the chromospheric heights along the ejected trajectories \citep{roy,sve,fou}.  Surges were studied for over seven decades and the different aspects of solar surge kinematics have been discussed by \cite{new,ell,gio,mac,kir}, and \cite{pla}.

Several authors, based on observational studies, have suggested that the driving mechanism of mass ejection in surges is magnetic reconnection at chromospheric or photospheric heights.  \cite{kur} showed that H$\alpha$ surges are seen in the early stages of flux emergence and suggested that these surges are produced by magnetic reconnection between newly emerged and the pre-existing magnetic flux.  These ideas are strongly supported by \cite{can} who reported that circumstances favorable to magnetic reconnection are produced by moving satellite spots in a surge-productive region.  Recently, \cite{udd} presented a multi-wavelength study of recurrent surges originated due to the photospheric reconnections.  \cite{can} also showed that a high-temperature X-ray jet and a cool untwisted surge can coexist located side by side at the site (see Fig.~10c in their paper). Although, \cite{tes} firstly reported that brief soft X-ray bursts are accompanied with solar surges.  From numerical MHD simulations, \cite{shi} and \cite{yok} succeeded to reproduce surge mass ejections from chromospheric heights by magnetic reconnection between an emerging flux and a pre-existing magnetic field.  This numerical result was later confirmed by \cite{nin} who studied the radio properties of $18$ X-ray coronal jets as observed by the \emph{Yohkoh\/} SXT \citep{tsu} using Nobeyama $17$~GHz data.  From the SXT images, \cite{nin} computed the coronal plasma parameters at the location of the surge.  At the time of maximum surge activity, they found electron temperature $T_{\rm e} = 2.8 \times 10^6$~K and emission measure $EM = 5.0 \times 10^{45}$~cm$^{-3}$, as well as derived constraints on the ejecta electron number density, notably $n_{\rm e} < 6.1 \times 10^{10}$~cm$^{-3}$.  H$\alpha$ surges and associated soft X-ray loops were also studied by \cite{sch} who performed simultaneous observations of NOAA AR 6850 on 1991 October 7, made with the MSDP spectrograph operating on the solar tower in Meudon and with the \emph{Yohkoh\/} SXT.  By measuring the volume emission measures of the two flaring loops (northern and southern ones) and the surge region (mid-part of the surge), \cite{sch} concluded that at 10:24--10:30 UT the temperature was ($3$--$4) \times 10^6$~K and the volume emission measure was $10^{47}$~cm$^{-3}$.  Assuming a volume of ($3$--$10) \times 10^{27}$~cm$^3$, they derived an electron number density $n_{\rm e} = (3$--$6) \times 10^9$~cm$^{-3}$.  \cite{kay} have observed a solar surge in NOAA AR 11271 using the \emph{Solar Dynamics Observatory\/} (\emph{SDO}) \citep{pes} data on 2011 August 25, possibly triggered by chromospheric activity.  They also measured the temperature and density distribution of the observed surge during its maximum rise and found an average temperature and a number density of $2 \times 10^6$~K and $4.17 \times 10^9$~cm$^{-3}$, respectively.  \cite{bri} studied the conditions for flares and surges in AR 2744 on 1980 October 21 and 22 using observations from the \emph{Solar Maximum Mission\/} satellite and coordinated ground-based observations, which together covered a wide temperature range from ${<}10^4$~K to ${>}10^7$~K.  In particular, the detected surge on October 22 had a total emission measure of $4.9 \times 10^{44}$~cm$^{-3}$ and duration of about $2000$~s.  The rough estimations of temperature and electron number density yielded $T_{\rm e} \sim 10^4$~K and $n_{\rm e} \sim 10^{12}$~cm$^{-3}$, respectively.

From a theoretical point of view, a surge can be modeled as a jet of high density (chromospheric density) flowing in magnetic flux tube.  As \cite{car} pointed out, such a jet is likely to be the site of instabilities which in turn trigger a nonlinear turbulent cascade with important effects on the solar corona and on the behavior of the jet itself.  It is well known that in a sheared velocity field, like that of surges, a Kelvin--Helmholtz (KH) instability may arise.  The KH instability was recently observed in the flaring coronal loop by \cite{cla}, as well as by \cite{ofm} who identified the formation, propagation, and decay of vortex-shaped features along the interface between an erupting (dimming) region and the surrounding corona imaged by the \emph{SDO}.  \cite{iva} have shown that the first imaged KH instability on coronal mass ejecta by \cite{clr} can be explained, in very good agreement with deduced observational data, as an instability of the $m = -3$ magnetohydrodynamic mode.  However, the evolution of such instabilities in surge-like flowing tubes where the significant flows affect the wave propagation, does not explored physically.  The aim of this study is to see whether magnetohydrodynamic (MHD) waves traveling along the surge jet can become unstable within its velocity range of $10$--$200$~km\,s$^{-1}$.  In the following Sec.~\ref{sec:model}, we will build up a simplified model for the surge.  Next Sec.~\ref{sec:dispeqn} deals with the derivation of the MHD wave dispersion relations, while in Sec.~\ref{sec:numerics} we will numerically analyze the dependence of the linear/threshold KH instability on relevant physical parameters of the surge and its environment.  The final, Sec.~\ref{sec:concl}, summarizes the results derived in this paper.

\section{Surge model and basic parameters}
\label{sec:model}
\begin{figure}[t]
   \centering
   \includegraphics[height=.30\textheight]{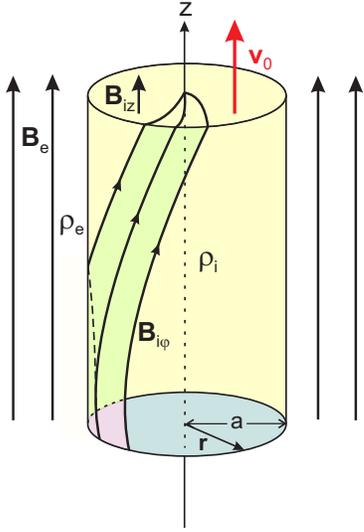}
   \caption{Equilibrium magnetic field geometry of an EUV solar surge. \emph{{Courtesy and Adaptation: R.~Erd\'elyi \& V.~Fedun 2010}}}
   \label{fig:fig1}
\end{figure}

We explore the high-temperature solar surge observed by \cite{kay} in NOAA AR11271 using the \emph{Solar Dynamics Observatory\/} data on 2011 August 25.  In addition to the temperature and electron number density inside the surge as already mentioned in the Introduction section, notably $2 \times 10^6$~K and $4.17 \times 10^9$~cm$^{-3}$, respectively, we state that surge's width is $\Delta \ell = 7$~Mm and it rises vertically from its origin up to a height of ${\approx}70$~Mm with a velocity of ${\approx}100$~km\,s$^{-1}$.  The deduced electron number density and temperature suggest that the main body of the surge is constituted with the plasma maintained at TR/inner coronal temperature.  Observational data yield also the electron number density in the environment, $n_{\rm e} = 1.15 \times 10^9$~cm$^{-3}$, and a lower electron temperature, $T_{\rm e} \approx 740\,000$~K, which can be taken to be roughly equal to $10^6$~K.  We assume that the magnetic field inside the jet is $10$~G, that implies an Alfv\'en speed $v_{\rm Ai} = 337.6$~km\,s$^{-1}$.  Under these circumstances, the pressure balance equation
\[
    p_{\rm i} + \frac{B_{\rm i}^2}{2\mu} = p_{\rm e} + \frac{B_{\rm e}^2}{2\mu},
\]
in which the subscript `i' implies \emph{interior\/} and `e' implies \emph{exterior},
requires the Alfv\'en speed outside the jet to be $v_{\rm Ae} \approx 719$~km\,s$^{-1}$, which means that the environment's magnetic field is $B_{\rm e} \approx 11.2$~G (more exactly, $11.1799$~G).  With sound speeds $c_{\rm i} = 166$~km\,s$^{-1}$ and
$c_{\rm e} = 117$~km\,s$^{-1}$, the plasma beta of both the media are respectively $\beta_{\rm i} = 0.3$ and $\beta_{\rm e} = 0.03$.  Thus, we can treat the surge as incompressible plasma and its environment as cool plasma.

We model the surge as a jet of homogeneous density $\rho_{\rm i}$ flowing with velocity $\bm{v}_0$ in a vertical magnetic flux tube of radius $a = \Delta \ell/2$ embedded in a magnetic field $\bm{B}_{\rm i}$ and surrounded by magnetized plasma of homogeneous density $\rho_{\rm e}$ with magnetic field $\bm{B}_{\rm e}$ (see Fig.~\ref{fig:fig1}).  We note that our frame of reference is attached
to the jet's environment, thus $\bm{v}_0$ must be considered as a relative velocity if there is any ambient flow.  In the general case, we assume that the equilibrium magnetic field inside the surge is twisted and the twist is characterized by the parameter $\varepsilon = B_{{\rm i}\varphi}/B_{{\rm i}z}$, where $B_{{\rm i}\varphi}$ is evaluated at the inner surface of the tube.  In cylindrical equilibrium, the magnetic field $\bm{B}_{\rm i}$ and thermal pressure $p_{\rm i}$ satisfy the equilibrium condition in the radial direction
\begin{equation}
\label{eq:equilib}
    \frac{\mathrm{d}}{\mathrm{d}r}\left( p_{\rm i} + \frac{B_{\rm i}^2}{2\mu}
    \right) = -\frac{B_{{\rm i} \varphi}^2}{\mu r},
\end{equation}
where, $B_{\rm i}(r) = \left( B_{{\rm i} \varphi}^2 + B_{{\rm i} z}^2 \right)^{1/2}$ denotes the strength of the equilibrium magnetic field, and $\mu$ is the magnetic permeability.  We note that in Eq.~(\ref{eq:equilib}) the total (thermal plus magnetic) pressure gradient is balanced by the tension force (the right-hand side of Eq.~(\ref{eq:equilib})) in the twisted field.  In our case of incompressible surge's plasma, we assume an equilibrium with uniform twist, i.e., the one for which $B_{{\rm i} \varphi}(r)/r B_{{\rm i} z}(r)$ is a constant.  Thus the background magnetic field is assumed to be
\begin{equation}
\label{eq:magnfield}
    \bm{B}_0(r) = \left\{ \begin{array}{lc}
                     (0, Ar, B_{{\rm i} z}) & \mbox{for $r \leqslant a$}, \\
                     (0, 0, B_{\rm e}) & \mbox{for $r > a$},
                              \end{array}
                      \right.
\end{equation}
where $A$, $B_{{\rm i} z}$, and $B_{\rm e}$ are constant. In the case of an untwisted surge, the twist parameter $\varepsilon = 0$ and magnetic field lines are straight lines as those of the environment's magnetic field $\bm{B}_{\rm e}$.

As seen from Fig.~\ref{fig:fig1}, except of the magnetic filed twist parameter $\varepsilon = Aa/B_{{\rm i}z}$, one can define the density contrast, $\eta = \rho_{\rm e}/\rho_{\rm i}$.  For an untwisted magnetic flux tube it is worth introducing the magnetic fields' parameter $b = B_{\rm e}/B_{\rm i}$.  For a twisted tube, this parameter takes the form $b = B_{\rm e}/B_{{\rm i}z}$.  Thus up to now, we have three input parameters for our surge model, notably $\varepsilon$, $b$, and $\eta$.

In studying the MHD wave propagation along an untwisted surge it will be significant to see how different will be the critical flow velocities for an evolved KH instability depending upon the choice of plasma beta: finite values of the sound speed (compressible plasmas) or consideration of very large sound speed inside the jet and its zero value in the environment.  Such a test will show that how acceptable is our assumption, especially in the case of twisted flux tube, to consider the surge's medium as incompressible plasma and its surrounding medium as cool magnetized plasma.

\section{Wave dispersion equations}
\label{sec:dispeqn}

In our system of cylindrical coordinates ($r, \varphi, z$), the equilibrium physical variables are functions of the radial coordinate $r$ only.  Then, the perturbed quantities can be Fourier-analyzed with respect to the ignorable coordinates $\varphi$ and $z$ and time $t$ and put proportional to $\exp[-\mathrm{i}(\omega t - m \varphi - k_z z)]$.  We can eliminate all but two of the perturbed variables (the perturbed total pressure $p_{\rm tot}$ and the radial component $\xi_r$ of the Lagrangian displacement ${\boldsymbol\xi}$) to get the following governing equations \cite{goo}:
\begin{equation}
    D \frac{\mathrm{d}}{\mathrm{d}r}\left( r \xi_r \right) = C_1 r \xi_r - C2 r p_{\rm tot},
\label{eq:xi}
\end{equation}
\begin{equation}
    D \frac{\mathrm{d}p_{\rm tot}}{\mathrm{d}r} = C_3 \xi_r - C_1 p_{\rm tot}.
\label{eq:ptot}
\end{equation}
The coefficient functions $D$, $C_1$, $C_2$, and $C_3$ depend on the equilibrium variables $\rho_0$, $\bm{B}_0$, $\bm{v}_0$ and on the Doppler-shifted frequency $\Omega = \omega - \bm{k}\cdot \bm{v}_0$.  In \cite{erd} notation, these coefficient functions are as follows:
\begin{equation}
    D = \rho_0 \left( \Omega^2 - \omega_{\rm A}^2 \right)C_4,
\label{eq:D}
\end{equation}
\begin{equation}
    C_1 = \frac{2B_{0\varphi}}{\mu r}\left( \Omega^4 B_{0\varphi} - \frac{m}{r}f_{\rm B}C_4 \right),
\label{eq:C1}
\end{equation}
\begin{equation}
    C_2 = \Omega^4 - \left( k_z^2 + \frac{m^2}{r^2} \right)C_4,
\label{eq:C2}
\end{equation}
\begin{eqnarray}
    C_3 = \rho_0 D \left[ \Omega^2 - \omega_{\rm A}^2 + \frac{2B_{0\varphi}}{\mu \rho_0} \frac{\mathrm{d}}{\mathrm{d}r} \left(  \frac{B_{0\varphi}}{r} \right) \right] \nonumber \\
    \nonumber \\
    {}+ 4\Omega^4 \left( \frac{B_{0\varphi}^2}{\mu r} \right)^2 - \rho_0 C_4 \frac{4B_{0\varphi}^2}{\mu r^2}\omega_{\rm A}^2,
\label{eq:C3}
\end{eqnarray}
where
\begin{equation}
    C_4 = \left( c_{\rm s}^2 + c_{\rm A}^2 \right)\left( \Omega^2 - \omega_{\rm c}^2 \right),
\label{eq:C4}
\end{equation}
and
\[
    f_{\rm B} = \frac{m}{r}B_{0\varphi} + \bm{k} \cdot \bm{B}_0, \;\; \omega_{\rm A}^2 = \frac{f_{\rm B}^2}{\mu \rho_0}, \;\; \omega_{\rm c}^2 = \frac{c_{\rm s}^2}{c_{\rm s}^2 + c_{\rm A}^2}\omega_{\rm A}^2.
\]
Here $\omega_{\rm A}$ is the Alfv\'en frequency and $\omega_{\rm c}$ is the cusp frequency; the other notation is standard.

After eliminating $\xi_r$, the set of Eqs.~(\ref{eq:xi}) and (\ref{eq:ptot}) can be rewritten in the form of a well known second-order ordinary differential equation \citep{hai,goe,sak}
\begin{eqnarray}
    \frac{\mathrm{d}^2 p_{\rm tot}}{\mathrm{d}r^2} + \left[ \frac{C_3}{rD} \frac{\mathrm{d}}{\mathrm{d}r} \left( \frac{rD}{C_3} \right) \right] \frac{\mathrm{d}p_{\rm tot}}{\mathrm{d}r} \nonumber \\
    \nonumber \\
    {}+ \left[ \frac{C_3}{rD} \frac{\mathrm{d}}{\mathrm{d}r} \left( \frac{rC_1}{C_3} \right) + \frac{1}{D^2} \left(  C_2 C_3 - C_1^2 \right) \right]p_{\rm tot} = 0.
\label{eq:ode}
\end{eqnarray}

Having obtained the solutions to Eq.~(\ref{eq:ode}) in both media, and finding the corresponding expressions for $\xi_r$, we can merge these solutions through appropriate boundary conditions at the interface $r = a$ and derive the dispersion relation of the normal modes propagating along the magnetic flux tube.

\subsection{Dispersion relation of MHD waves on an untwisted flux tube}
\label{subsec:untwist}
For an untwisted magnetic flux tube the coefficients $C_1 = 0$ and $C_3 = \rho_0 D \left( \Omega^2 - \omega_{\rm A}^2 \right)$; then Eq.~(\ref{eq:ode}) takes the form
\begin{equation}
    \frac{\mathrm{d}^2 p_{\rm tot}}{\mathrm{d}r^2} + \frac{1}{r}\frac{\mathrm{d}p_{\rm tot}}{\mathrm{d}r} - \left( m_0^2 + \frac{m^2}{r^2} \right)p_{\rm tot} = 0,
\label{eq:besssel}
\end{equation}
where
\begin{equation}
    m_0^2 = -\frac{\left( \Omega^2 - k_z^2 c_{\rm s}^2 \right)\left( \Omega^2 - k_z^2 v_{\rm A}^2 \right)}{\left( c_{\rm s}^2 + v_{\rm A}^2 \right)\left( \Omega^2 - \omega_{\rm c}^2 \right)}.
\label{eq:m0sqr}
\end{equation}
The cusp frequency, $\omega_{\rm c}$, is usually expressed via the so called tube speed, $c_{\rm T}$, notably $\omega_{\rm c} = k_z  c_{\rm T}$, where \citep{edw}
\[
    c_{\rm T} = \frac{c_{\rm s}v_{\rm A}}{\sqrt{c_{\rm s}^2 + v_{\rm A}^2}}.
\]
The solutions for $p_{\rm tot}$ can be written in terms of modified Bessel functions: $I_m(m_{0{\rm i}}r)$ inside the jet and $K_m(m_{0{\rm e}}r)$ in its environment.  The wave attenuation coefficients, $m_{0{\rm i}}$ and $m_{0{\rm e}}$, in both media are calculated from Eq.~(\ref{eq:m0sqr}) with replacing the sound and Alfv\'en speeds with the corresponding values for each medium.  Note that in evaluating $m_{0{\rm e}}$, the wave frequency is simply $\omega$.  By expressing the Lagrangian displacements in both media via the derivatives of corresponding Bessel functions and applying the boundary conditions for continuity of $p_{\rm tot}$ and $\xi_r$ across the interface, $r = a$, one obtains the dispersion relation of normal MHD modes propagating in a flowing compressible jet surrounded by a static compressible plasma \citep{ter,nak,zhe}
\begin{eqnarray}
\label{eq:dispeq}
	\frac{\rho_{\rm e}}{\rho_{\rm i}}\left( \omega^2 - k_z^2 v_{\rm Ae}^2
        \right) m_{0{\rm i}}\frac{I_m^{\prime}(m_{0{\rm
        i}}a)}{I_m(m_{0{\rm i}}a)} \nonumber \\
        \nonumber \\
        {}- \left[ \left( \omega - \bm{k} \cdot
        \bm{v}_0 \right)^2 - k_z^2 v_{\rm Ai}^2 \right] m_{0{\rm
        e}}\frac{K_m^{\prime}(m_{0{\rm e}}a)}{K_m(m_{0{\rm e}}a)} = 0.
\end{eqnarray}
It is clearly seen that the wave frequency is Doppler-shifted inside the jet.  We recall that for the kink mode ($m = 1$) one defines the kink speed \citep{edw}
\[
	c_{\rm k} = \left( \frac{\rho_{\rm i} v_{\rm Ai}^2 + \rho_{\rm e}
        v_{\rm Ae}^2}{\rho_{\rm i} + \rho_{\rm e}} \right)^{1/2} = \left(
        \frac{v_{\rm Ai}^2 + (\rho_{\rm e}/\rho_{\rm i})v_{\rm Ae}^2}{1 +
        \rho_{\rm e}/\rho_{\rm i}} \right)^{1/2},
\]
which is independent of sound speeds and characterizes the propagation of
transverse perturbations.  As we will demonstrate, the kink mode can become unstable against the KH instability.

In the case of surge's incompressible plasma and cool one for its environment, the MHD wave dispersion relation (\ref{eq:dispeq}) keeps its form, but the two attenuation coefficients $m_{0{\rm i,e}}$ become much simpler, notably
\[
    m_{0{\rm i}} = k_z
\]
and
\[
    m_{0{\rm e}} = \left( k_z^2 v_{\rm Ae}^2  - \omega^2 \right)^{1/2}/v_{\rm Ae},
\]
respectively.

\subsection{Dispersion relation of MHD waves on a twisted flux tube}
\label{subsec:twist}
Inside the flux tube ($r \leqslant a$), with $B_{{\rm i}\varphi} = Ar$, the coefficient $f_{\rm B}$ and the Alfv\'en frequency $\omega_{\rm Ai}$ take the forms
\[
    f_{\rm B} = mA + k_z B_{{\rm i}z} \quad \mbox{and} \quad \omega_{\rm Ai} = \frac{mA + k_z B_{{\rm i}z}}{\sqrt{\mu \rho_{\rm i}}},
\]
respectively.  Following \cite{erd}, we redefine the coefficient functions $D$, $C_1$, $C_2$, and $C_3$ for incompressible plasma by dividing them by $C_4$ to get
\[
    D = \rho \left( \Omega^2 - \omega_{\rm A}^2 \right), \quad C_1 = -\frac{2mB_{\varphi}}{\mu r^2}\left( \frac{m}{r}B_{\varphi} + k_z B_z \right),
\]
\[
     C_2 = - \left( \frac{m^2}{r^2} + k_z^2 \right),
\]
\[
     C_3 = D^2 + D\frac{2 B_{\varphi}}{\mu}\frac{\mathrm{d}}{\mathrm{d} r} \left( \frac{B_{\varphi}}{r} \right) - \frac{4 B_{\varphi}^2}{\mu r^2}\rho \omega_{\rm A}^2.
\]
Radial displacement $\xi_{r}$ is expressed through the total pressure perturbation as
\begin{equation}
\label{eq:displ}
    \xi_{r} = \frac{D}{C_3}\frac{\mathrm{d} p_{\rm tot}}{\mathrm{d} r} + \frac{C_1}{C_3}p_{\rm tot}.
\end{equation}
The solution to this equation depends on the magnetic field and density profile.

With aforementioned coefficient functions $D$, $C_1$, $C_2$, and $C_3$, written down for the interior, Eq.~(\ref{eq:ode}) reduces to the modified Bessel equation
\begin{equation}
\label{eq:diffeq}
	\left[ \frac{\mathrm{d}^2}{\mathrm{d}r^2} + \frac{1}{r}
        \frac{\mathrm{d}}{\mathrm{d} r} - \left( m_{\rm 0i}^2 + \frac{m^2}{r^2}
        \right) \right] p_{\rm tot} = 0,
\end{equation}
where
\begin{equation}
\label{eq:kappa}
	m_{\rm 0i}^2 = k_z^2 \left[  1 - \frac{4 A^2 \omega_{\rm Ai}^2}
        {\mu \rho_{\rm i} \left( \Omega^2 -
        \omega_{\rm Ai}^2\right)^2} \right].
\end{equation}

The solution to Eq.~(\ref{eq:diffeq}) bounded at the tube axis is
\begin{equation}
\label{eq:solution}
    p_{\rm tot}(r \leqslant a) = \alpha_{\rm i}I_m(m_{\rm 0i}r),
\end{equation}
where $I_m$ is the modified Bessel function of order $m$ and $\alpha_{\rm i}$ is a constant.  Transverse displacement $\xi_{{\rm i}r}$ can be written using Eq.~(\ref{eq:displ}) as
\begin{eqnarray}
\label{eq:xir}
    \xi_{{\rm i}r} = \frac{\alpha_{\rm i}}{r}\left\{ \frac{\left( \Omega^2 - \omega_{\rm Ai}^2 \right)m_{\rm 0i}r I_m^{\prime}(m_{\rm 0i}r)}{\rho_{\rm i}\left( \Omega^2 - \omega_{\rm Ai}^2 \right)^2 - 4A^2 \omega_{\rm Ai}^2/\mu} \right. \nonumber \\
    \nonumber \\
    \left.
     - \frac{2mA\omega_{\rm Ai}I_m(m_{\rm 0i}r)/\sqrt{\mu \rho_{\rm i}} }{\rho_{\rm i}\left( \Omega^2 - \omega_{\rm Ai}^2 \right)^2 - 4A^2 \omega_{\rm Ai}^2/\mu} \right\},
\end{eqnarray}
where the prime sign means a differentiation by the Bessel function argument.

For the cool environment with straight-line magnetic field $B_{{\rm e}z} = B_{\rm e}$ and homogeneous density $\rho_{\rm e}$, the $C_{1\text{--}3}$ and $D$ coefficients have the form
\[
    D = \rho \left( \omega^2 - \omega_{\rm A}^2 \right), \quad C_1 = 0,
\]
\[
    C_2 = -\left[ \frac{m^2}{r^2} + k_z^2 \left( 1 - \frac{\omega^2}{\omega_{\rm A}^2}\right) \right], \quad C_3 = D^2.
\]
The total pressure perturbation outside the tube is governed by the same Bessel equation as Eq.~(\ref{eq:diffeq}), but $m_{\rm 0i}^2$ is replaced by
\begin{equation}
\label{eq:kappae}
    m_{\rm 0e}^2 = k_z^2 \left( 1 - \omega^2/\omega_{\rm Ae}^2 \right),
\end{equation}
which is just the same as the attenuation coefficient in the cool environment of an untwisted flux tube.  The solution bounded at infinity is
\begin{equation}
\label{eq:pmag}
    p_{\rm tot}(r > a) = \alpha_{\rm e}K_m(m_{\rm 0e}r),
\end{equation}
where $K_m$ is the modified Bessel function of order $m$ and $\alpha_{\rm e}$ is a constant.

The transverse displacement now can be written as
\begin{equation}
\label{eq:xioutside}
    \xi_{{\rm e}r} = \frac{\alpha_{\rm e}}{r}\frac{m_{\rm 0e}rK_m^{\prime}(m_{\rm 0e}r)}{\rho_{\rm e}\left( \omega^2 - \omega_{\rm Ae}^2 \right)},
\end{equation}
where the prime sign means as before a differentiation by the Bessel function argument and
\begin{equation}
\label{eq:omegaae}
    \omega_{\rm Ae} = \frac{k_z B_{{\rm e}z}}{\sqrt{\mu \rho_{\rm e}}} = k_z v_{\rm Ae}.
\end{equation}
Here, $v_{\rm Ae} = B_{\rm e}/\sqrt{\mu \rho_{\rm e}}$ is the Alfv\'en speed in the surrounding magnetizes plasma.

The boundary conditions which merge the solutions inside and outside the twisted magnetic flux tube are the continuity of the radial component of the Lagrangian displacement
\[
    \xi_{{\rm i}r}|_{r=a} = \xi_{{\rm e}r}|_{r=a}
\]
and the total pressure perturbation \citep{ben}
\[
    \left.p_{\rm tot\,i} - \frac{B_{{\rm i}\varphi}^2}{\mu a}\xi_{{\rm i}r}\right\vert_{r=a} = p_{\rm tot\,e}|_{r=a},
\]
where total pressure perturbations $p_{\rm tot\,i}$ and $p_{\rm tot\,e}$ are given by Eqs.~(\ref{eq:solution}) and (\ref{eq:pmag}), respectively.
Applying these boundary condition, after some algebra we finally derive the dispersion relation of the normal MHD modes propagating along a twisted magnetic flux tube with axial mass flow $\bm{v}_0$
\begin{eqnarray}
\label{eq:twdispeq}
	\frac{\left( \Omega^2 -
    \omega_{\rm Ai}^2 \right)F_m(m_{\rm 0i}a) - 2mA \omega_{\rm Ai}/\sqrt{\mu \rho_{\rm i}}}
    {\left( \omega^2 -
    \omega_{\rm Ai}^2 \right)^2 - 4A^2\omega_{\rm Ai}^2/\mu \rho_{\rm i} } \nonumber \\
    \nonumber \\
    {} = \frac{P_m(m_{\rm 0e} a)}
    {{\displaystyle \frac{\rho_{\rm e}}{\rho_{\rm i}}} \left( \omega^2 - \omega_{\rm Ae}^2
    \right) + A^2  P_m(m_{\rm 0e} a)/\mu \rho_{\rm i}},
\end{eqnarray}
where, remember, $\Omega = \omega - \bm{k}\cdot \bm{v}_0$ is the Doppler-shifted wave frequency in the moving medium,
\[
    F_m(m_{\rm 0i}a) = \frac{m_{\rm 0i}a I_m^{\prime}(m_{\rm 0i}a)}{I_m(m_{\rm 0i}a)},
\]
and
\[
    P_m(m_{\rm 0e}a) = \frac{m_{\rm 0e}a K_m^{\prime}(m_{\rm 0e}a)}{K_m(m_{\rm 0e}a)}.
\]
This dispersion equation is similar to the dispersion equation of normal MHD modes in a twisted flux tube surrounded by incompressible plasma \citep{izh} -- there, in Eq.~(13), $\kappa_{\rm e} \equiv m_{\rm 0e} = k_z$, and to the dispersion equation for a twisted tube with non-magnetized environment, i.e., with $\omega_{\rm Ae} = 0$ \citep{tza}.

\section{Numerical calculations and results}
\label{sec:numerics}

The main goal of our study is to see under which conditions the propagating  MHD waves along the jet can become unstable.  To conduct such an investigation it is necessary to assume that the wave frequency $\omega$ is a complex quantity, i.e., $\omega \to \omega + \mathrm{i}\gamma$, where $\gamma$ is the instability growth rate, while the longitudinal wavenumber $k_z$ is a real variable in the wave dispersion relation.  Since the occurrence of the expected KH instability is determined primarily by the jet velocity, therefore, to obtain a critical/threshold value of it, we will gradually change its magnitude from zero to that critical value and beyond.  Thus, we have to solve the dispersion relations in complex variables obtaining the real and imaginary parts of the wave frequency, or as it is normally accepted, of the wave phase velocity $v_{\rm ph} = \omega/k_z$, as functions of $k_z$ at various values of the velocity shear between the surge and its environment, $v_0$.

We first begin with the numerical solving Eq.~(\ref{eq:dispeq}) assuming that both the surge and its environment are compressible plasmas.  Since for this equation the magnetic flux tube is untwisted, the twist parameter $\varepsilon = 0$.  Before starting the numerical job, we have to normalize all variables and to specified the input parameters.  The wave phase velocity, $v_{\rm ph}$, and the other speeds are normalized to the Alfv\'en speed inside the jet, $v_{\rm Ai}$, which is calculated on using the axial magnetic fields $B_{\rm i}$ (for untwisted tube) or $B_{{\rm i}z}$ (for twisted surge).  The wavelength, $\lambda = 2\pi/k_z$, is normalized to the tube radius, $a$, that is equivalent to introducing a dimensionless wavenumber $k_z a$.  For compressible plasmas, except the density contrast $\eta = 0.28$ and magnetic fields ratio $b = 1.118$, we have to additionally specify other two input parameters, notably the ratio $\tilde{\beta} = c_{\rm s}^2/v_{\rm A}^2$ for the two media.  In accordance with the discussed in Sec.~\ref{sec:model} sound and Alfv\'en speeds, we have $\tilde{\beta}_{\rm i} = 0.2416$ and $\tilde{\beta}_{\rm e} = 0.0266$, respectively.  In the dimensionless analysis the flow speed, $v_0$, will be presented by the Alfv\'en Mach number $M_{\rm A} = v_0/v_{\rm Ai}$.

\begin{figure}[!ht]
\centering
    \includegraphics[width=8.0cm]{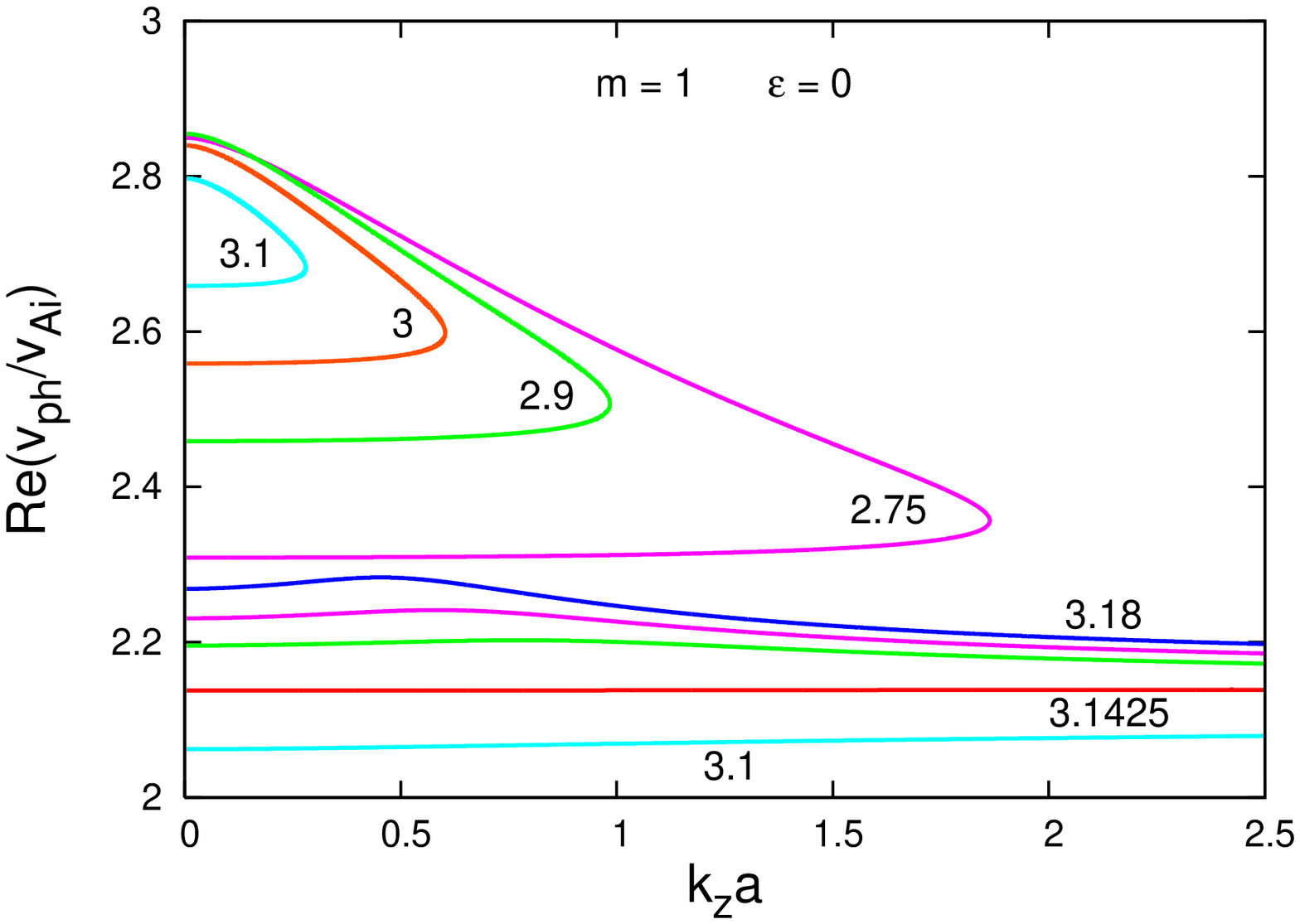} \\
\vspace{1mm}
    \includegraphics[width=8.0cm]{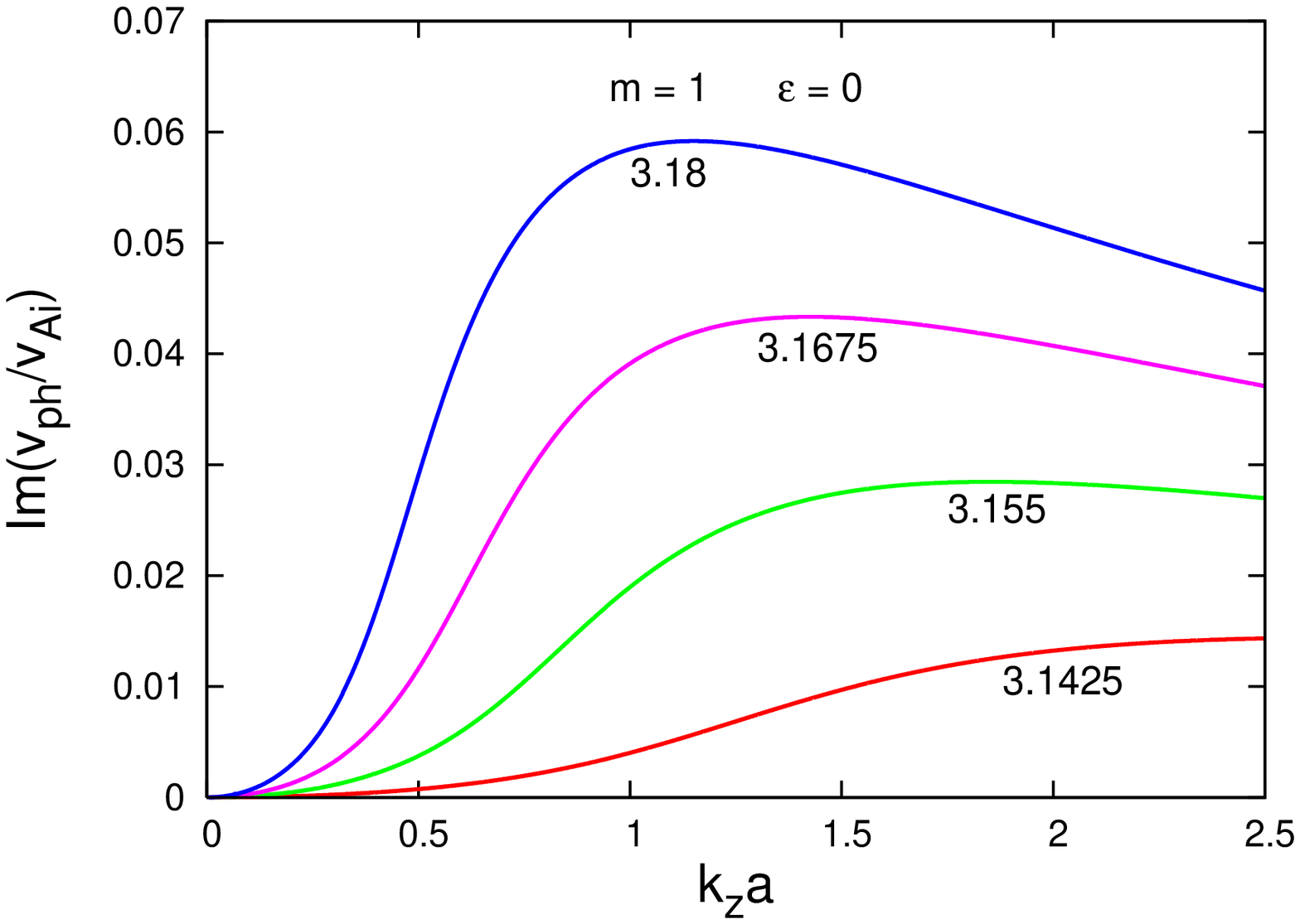}
   \caption{(\emph{Top panel}) Dispersion curves of stable and unstable kink ($m = 1$) modes propagating on a jet in an untwisted external magnetic field ($\varepsilon = 0$) at $\eta = 0.28$, $b = 1.118$, $\tilde{\beta}_{\rm i} = 0.2416$, $\tilde{\beta}_{\rm e} = 0.0266$, and various Alfv\'en Mach numbers.  (\emph{Bottom panel}) Growth rates of the unstable kink mode for Alfv\'en Mach numbers equal to $3.1425$, $3.155$, $3.1675$, and $3.18$, respectively}
   \label{fig:fig2}
\end{figure}

Among the wave modes which one can `extract' from dispersion Eq.~(\ref{eq:dispeq}), we will be interested in the kink, $m = 1$, mode because, as previous studies (Andries \& Goossens 2001; Vasheghani Farahani et al.\ 2009; Zhelyazkov 2012; Zhelyazkov \& Zaqarashvili 2012; Zaqarashvili et al.\ 2014) show, namely the kink mode becomes unstable when the jet speed exceeds a critical value.  At a static magnetic flux tube ($M_{\rm A} = 0$), the kink mode propagates with the kink speed, $c_{\rm k}$, which in our case is equal to $448.3$~km\,s$^{-1}$, or in dimensionless units, $1.33$, i.e., the wave is slightly super-Alfv\'enic.  The flow shifts upwards the kink-speed dispersion curve as well as splits it into two separate curves, which for small Alfv\'en Mach numbers travel with speeds $M_{\rm A} \mp c_{\rm k}/v_{\rm Ai}$ \citep{zhe}.  (A similar duplication happens for the tube-speed dispersion curves, too.)  At higher $M_{\rm A}$, however, the behavior of each curve of the pair $M_{\rm A} \mp c_{\rm k}/v_{\rm Ai}$ is completely different.  As seen from Fig.~\ref{fig:fig2}, for $M_{\rm A} \geqslant 2.75$ the higher kink-speed
wave, at some dimensionless wavenumber $k_z a$, makes a turn forming a semi-closed curve, while the lower kink-speed wave propagates with practically constant phase speed -- this can be seen in the top panel of Fig.~\ref{fig:fig2} for the $M_{\rm A} = 3.1$ pair of dispersion curves.  The kink mode becomes unstable at $M_{\rm A}^{\rm cr} = 3.1425$ and this is the lower kink-speed wave; the higher kink-speed wave is stable one and has the shape of a semi-closed curve (not plotted on the wave dispersion diagram).  The growth rates of the unstable kink mode are plotted in the bottom panel of Fig.~\ref{fig:fig2}.  The red curves in all diagrams denote the marginal dispersion/growth rate curves: for $M_{\rm A} < M_{\rm A}^{\rm cr}$ the kink mode is stable, otherwise it is unstable and the instability is of the KH type.  With $M_{\rm A}^{\rm cr} = 3.1425$ the lower kink-speed wave will be unstable if the jet velocity exceeds $1060$~km\,s$^{-1}$ -- a speed, which is far beyond the speed accessible for solar surges.
\begin{figure}[!ht]
\centering
    \includegraphics[width=8.0cm]{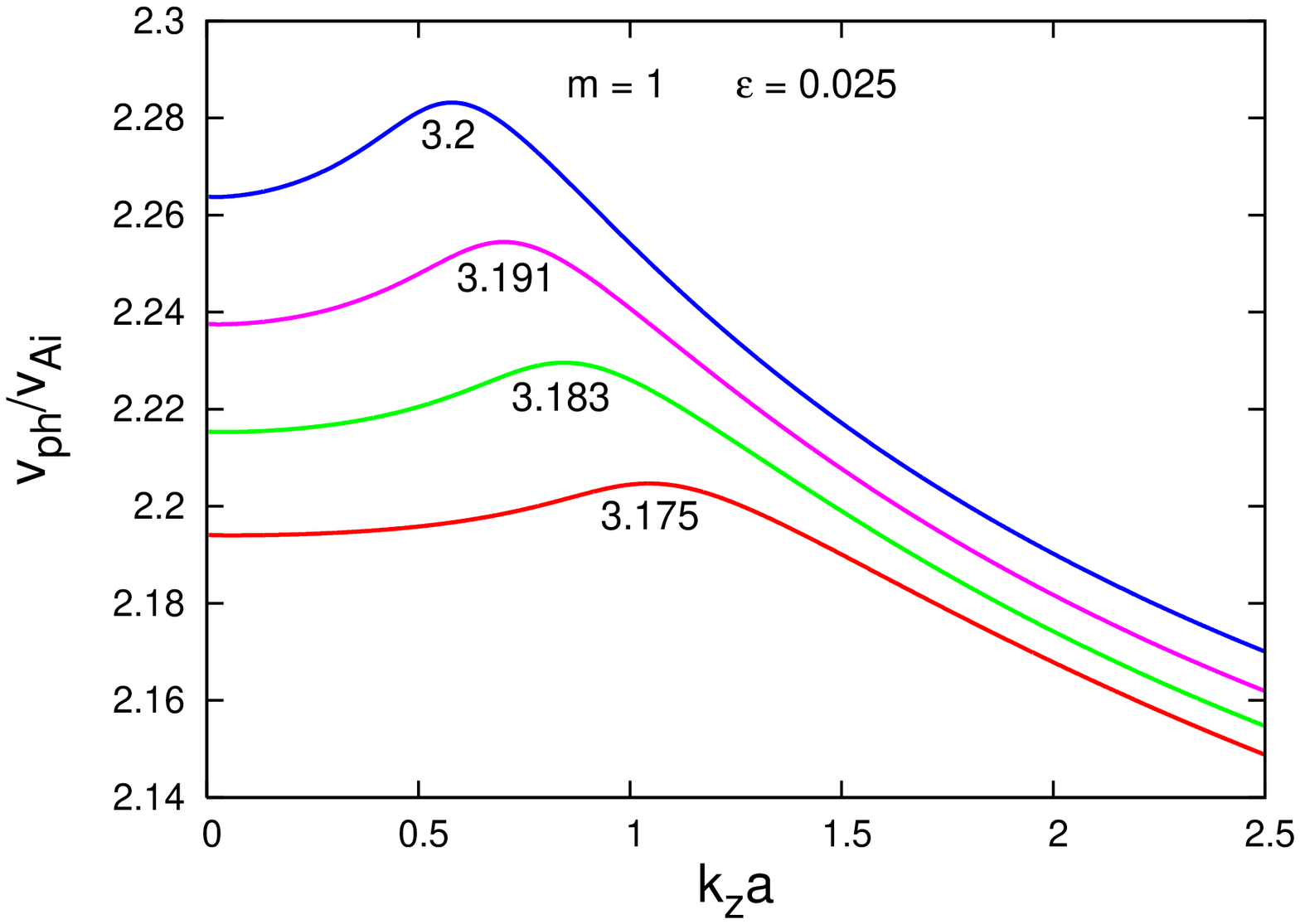} \\
\vspace{1mm}
    \includegraphics[width=8.0cm]{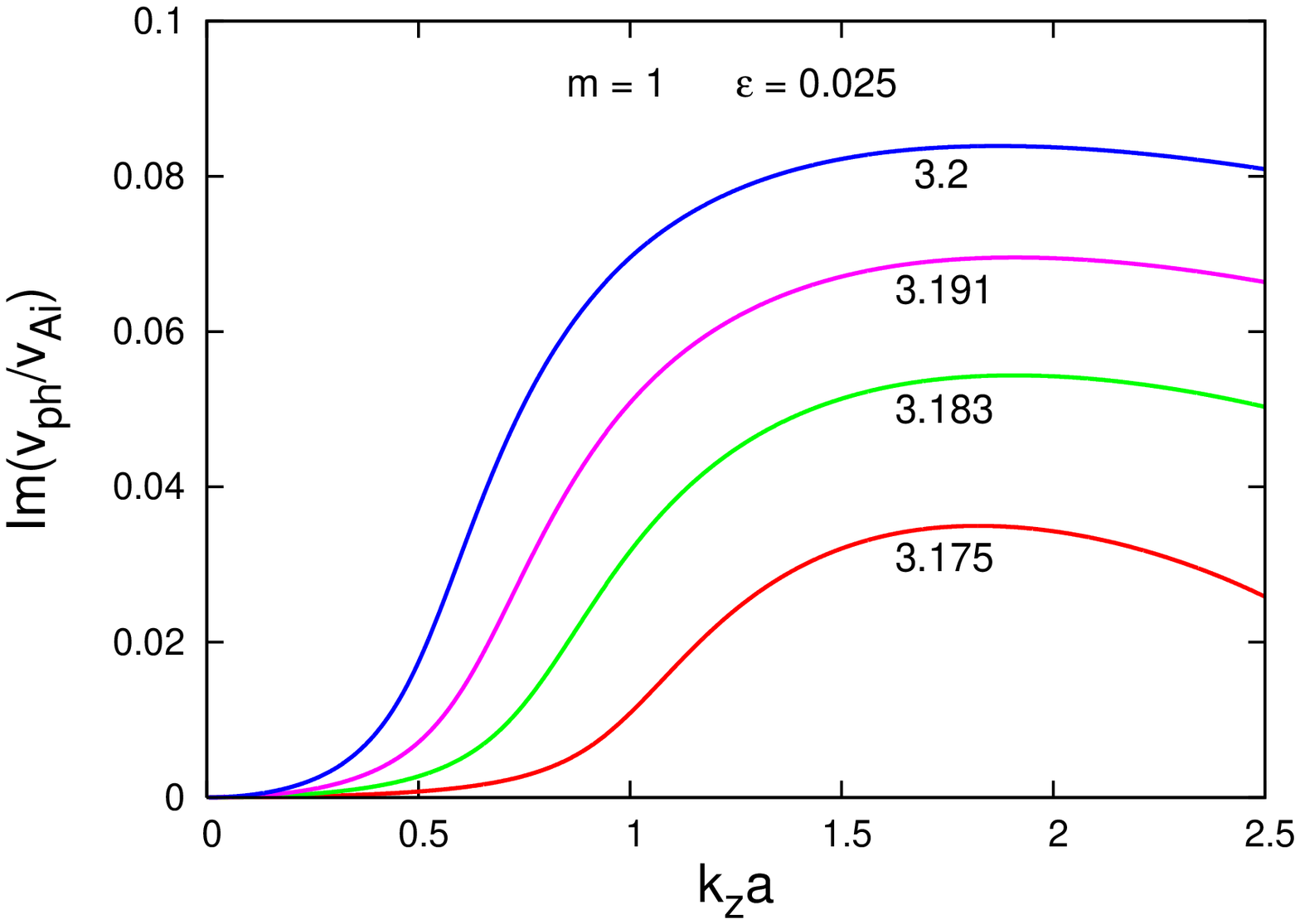} \\
   \caption{(\emph{Top panel}) Dispersion curves of unstable kink ($m = 1$) mode propagating on an incompressible jet in a twisted internal magnetic field ($\varepsilon = 0.025$) at $\eta = 0.28$, $b_{\rm m} = 1.1357$, and various Alfv\'en Mach numbers.  (\emph{Bottom panel}) Growth rates of unstable kink mode for Alfv\'en Mach numbers equal to $3.175$, $3.183$, $3.191$, and $3.2$, respectively}
   \label{fig:fig3}
\end{figure}
Thus, the kink wave traveling along the jet is stable against the KH instability.  It is important to note that the unstable dispersion curves and growth rates derived from the simplified form of dispersion equation (\ref{eq:dispeq}) (incompressible jet and cool environment) are very similar to those plotted in Fig.~\ref{fig:fig2} (we note that the assumption $\beta_{\rm e} = 0$ requires a slightly higher external magnetic field, $b = 1.1357$, and correspondingly a new Alfv\'en speed $v_{\rm Ae} = 730$~km\,s$^{-1}$).  The critical Alfv\'en Mach number now is equal to $3.17$ -- it is not surprising that this value is slightly greater than $3.1425$ because it is well established that the compressibility enhances the threshold Alfv\'en Mach numbers.  Anyway, the comparison shows that our surge's model of flowing incompressible plasma surrounded by a cool environment is reliable and can be used in studying the MHD waves propagation in the more complex case of a twisted magnetic flux tube.
\begin{figure}[!ht]
\centering
    \includegraphics[width=8.0cm]{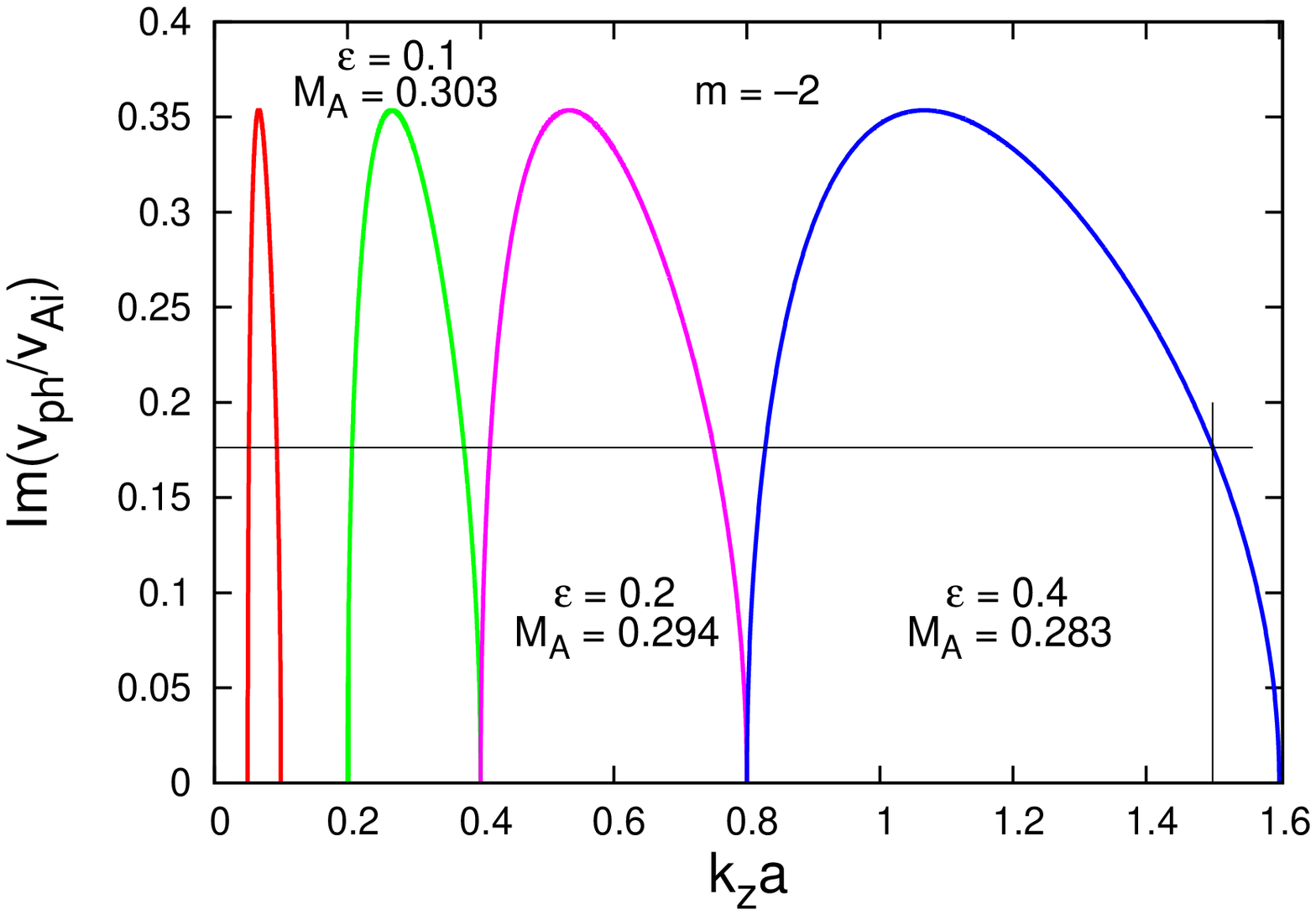} \\
\vspace{1mm}
    \includegraphics[width=8.0cm]{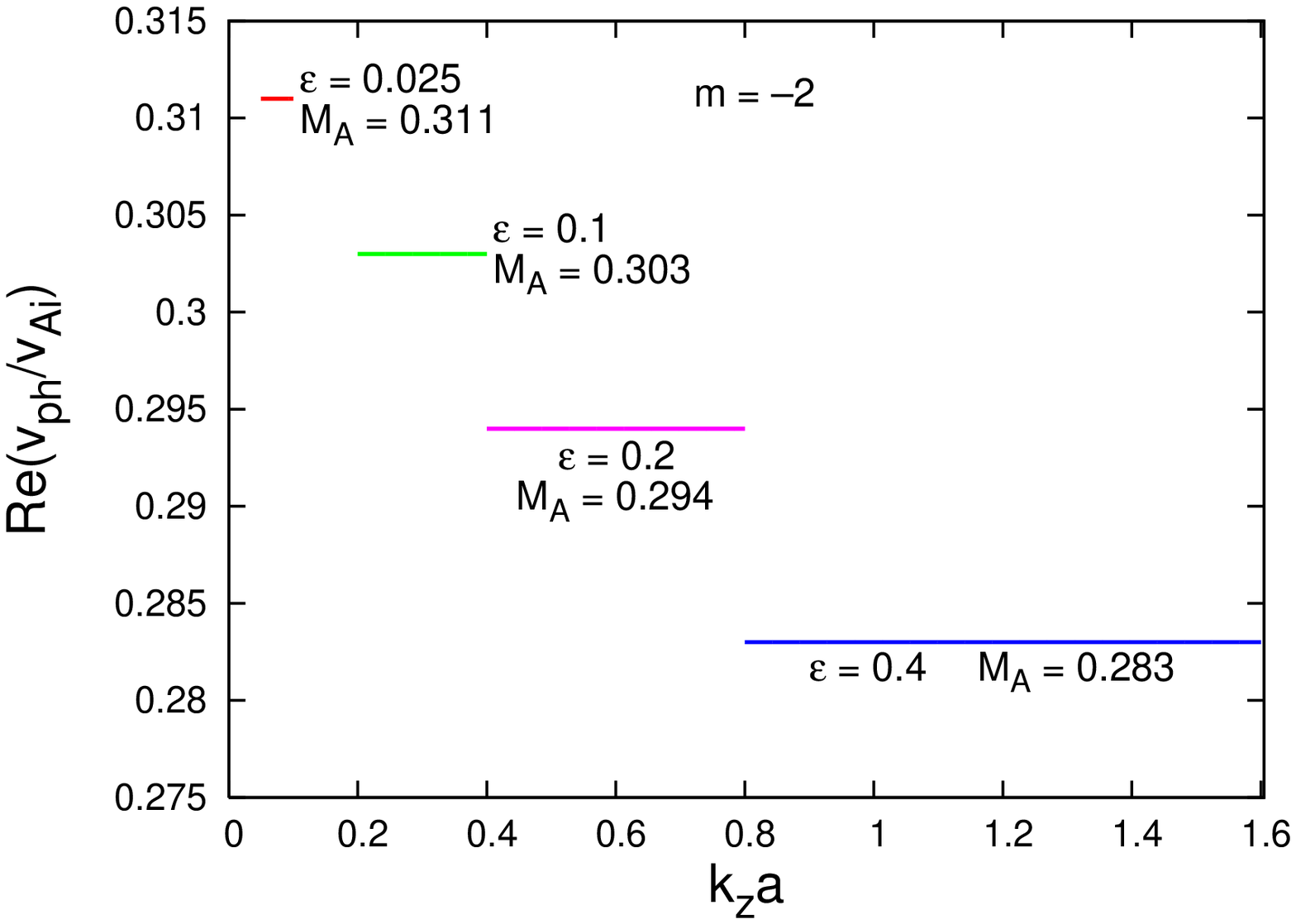} \\
   \caption{(\emph{Top panel}) Growth rates of the unstable $m = -2$ MHD mode propagating on incompressible jets in four different twisted internal magnetic fields (with $\varepsilon = 0.025$, $0.1$, $0.2$, and $0.4$) at $\eta = 0.28$, $b = 1.1357$, and corresponding critical Alfv\'en Mach numbers.  For $k_za = 1.5$ the wavelength of the unstable $m = -2$ harmonic is $\lambda_{\rm KH} = 14.7$~Mm, and the wave growth rate is $\gamma_{\rm KH} = 0.026$~s$^{-1}$.  (\emph{Bottom panel}) Marginal dispersion curves of the unstable $m = -2$ MHD mode for the critical Alfv\'en Mach numbers as functions of the magnetic field twist parameter $\varepsilon$}
   \label{fig:fig4}
\end{figure}
\begin{figure}[!ht]
\centering
    \includegraphics[width=8.0cm]{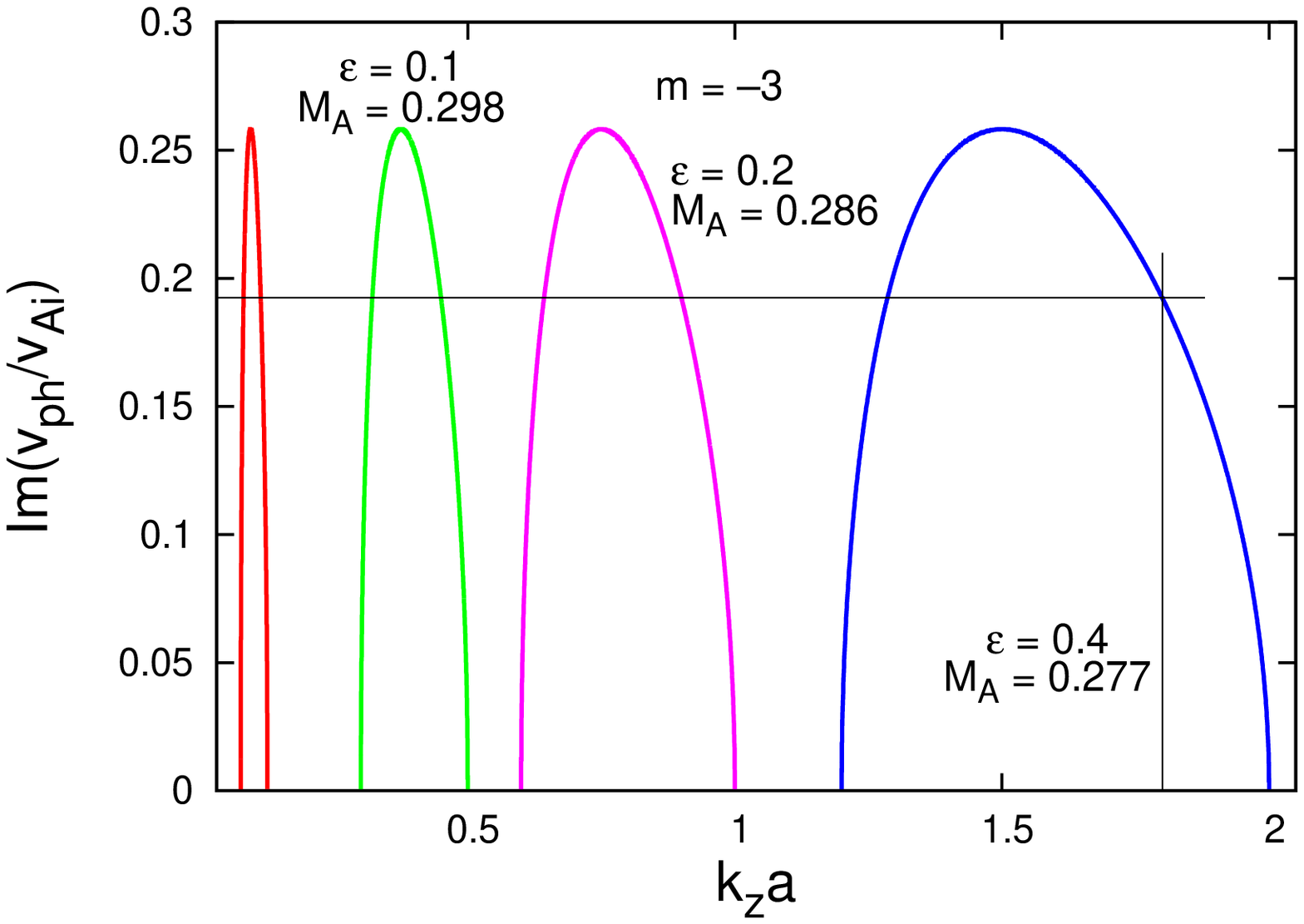} \\

\vspace{1mm}
    \includegraphics[width=8.0cm]{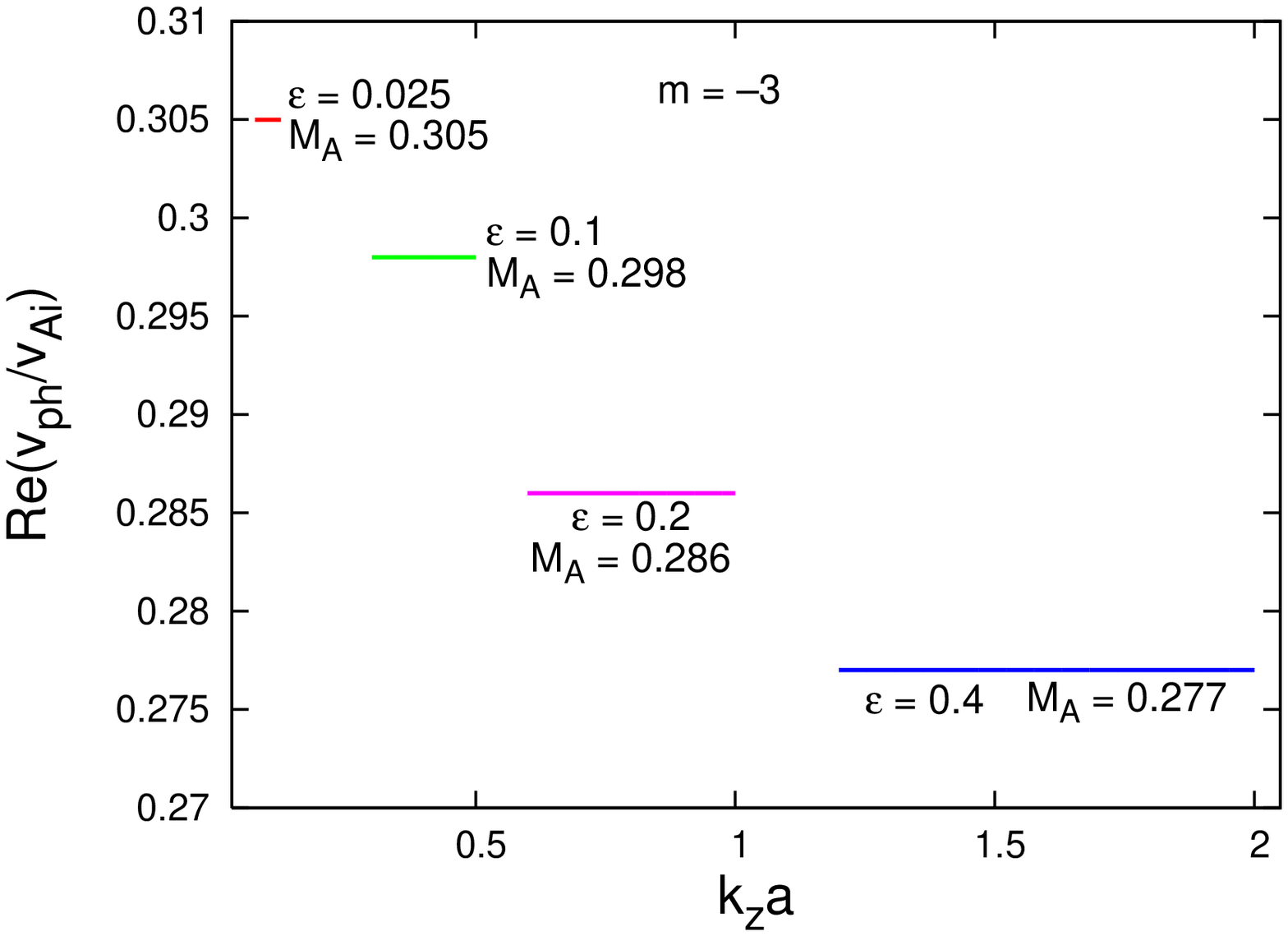} \\
   \caption{(\emph{Top panel}) Growth rates of the unstable $m = -3$ MHD mode propagating on incompressible jets in four different twisted internal magnetic fields (with $\varepsilon = 0.025$, $0.1$, $0.2$, and $0.2$) at $\eta = 0.28$, $b = 1.1357$, and corresponding critical Alfv\'en Mach numbers.  For $k_za = 1.8$ the wavelength of the unstable $m = -3$ harmonic is $\lambda_{\rm KH} = 12.2$~Mm, and the wave growth rate is $\gamma_{\rm KH} = 0.033$~s$^{-1}$.  (\emph{Bottom panel}) Marginal dispersion curves of the unstable $m = -3$ MHD mode for the critical Alfv\'en Mach numbers as functions of the magnetic field twist parameter $\varepsilon$}
   \label{fig:fig5}
\end{figure}

In a twisted magnetic flux tube the shape of dispersion curves and growth rates of the unstable kink, $m = 1$, mode is similar to that in an untwisted tube.  This is seen in Fig.~\ref{fig:fig3} where we display such curves derived from the numerical solving Eq.~(\ref{eq:twdispeq}) at $\varepsilon = 0.025$, $\eta = 0.28$, and $b = 1.1357$.  The most surprising result is the circumstance that the critical Alfv\'en Mach number for the KH instability rising (equal to $3.175$) is slightly higher than that for an untwisted tube (equal to $3.17$) at the same approximation (incompressible jet and cool environment).  A similar comparison between the threshold Alfv\'en Mach numbers in twisted and untwisted flux tubes for isolated photospheric jets treated as incompressible plasmas is just the opposite: the magnetic field twist decreases the critical numbers/velocities for an instability emergence \citep{izh}.  An increase in the twist parameter, $\varepsilon$, to $0.1$ or $0.2$ does not change significantly the wave dispersion characteristics and critical Alfv\'en Mach numbers -- as in the case of untwisted flux tube a critical jet speed of $1072$~km\,s$^{-1}$ is not accessible for solar surges.

A distinctive decrease of the instability critical Alfv\'en Mach number/jet speed one can achieve at the propagation of higher MHD modes.  As it was shown by \cite{tza}, then one can have even sub-Alfv\'enic critical jet speeds.  We have studied that issue for the mode numbers $|m| = 2$ and $|m| = 3$.  Calculations show that the $m = 2$ and $m = 3$ harmonics are stable against the KH instability -- one cannot obtain unstable solutions for reasonable Alfv\'en Mach numbers.  For the $m = -2$ and $m = -3$ MHD modes, however, one can.  Numerical calculations of Eq.~(\ref{eq:twdispeq}) for the $m = -2$ MHD mode yield four instability windows on the $k_za$-axis whose position and width depends upon the magnetic field twist parameter $\varepsilon$ (see Fig.~\ref{fig:fig4}).  The critical jet speeds for emergence of a KH instability accordingly are $105$~km\,s$^{-1}$, $102.3$~km\,s$^{-1}$, $99.3$~km\,s$^{-1}$, and $95.5$~km\,s$^{-1}$.  The two narrow windows corresponding to $\varepsilon = 0.025$ and $\varepsilon = 0.1$ are practically inapplicable to our surge--environment configuration: the wavelength, $\lambda = \pi \Delta \ell/k_z a$, of unstable $m = -2$ harmonics is comparable to the surge's height, ${\approx}70$~Mm.  Actually only in the forth instability window (at $\varepsilon = 0.4$) one can have real unstable $m = -2$ MHD mode: for instance, at $k_za = 1.5$ the wavelength is $\lambda_{\rm KH} = 14.7$~Mm, and the corresponding wave growth rate is $\gamma_{\rm KH} = 0.026$~s$^{-1}$.  For the $m = -3$ harmonic one obtains also four instability windows (see Fig.~\ref{fig:fig5}) with different positions and widths compared to those of the unstable $m = -2$ MHD mode.  The critical surge's speeds now are $103$~km\,s$^{-1}$, $100.6$~km\,s$^{-1}$, $96.6$~km\,s$^{-1}$, and $93.5$~km\,s$^{-1}$, respectively.  A simple evaluation shows that again in the forth instability window one can detect unstable $m = -3$ harmonics.  If we pick out $k_za = 1.8$, the wavelength of the unstable mode is $\lambda_{\rm KH} = 12.2$~Mm with a growth rate $\gamma_{\rm KH} = 0.033$~s$^{-1}$.  It is worth underlying that the magnitude of $\gamma_{\rm KH}$ for each unstable MHD mode crucially depends on the $k_za$-value: in the forth instability window of the $m = -2$ MHD mode at $k_za \approx 1.067$ (the maximum value of Im$(v_{\rm ph}/v_{\rm Ai})$) the wave growth rate becomes equal to $0.036$~s$^{-1}$; for the $m = -3$ harmonic at $k_z a = 1.5$ it is $0.037$~s$^{-1}$.  It is common for both modes that the unstable wave phase velocities coincide with the jet speeds (one sees in the bottom panels of Figs.~\ref{fig:fig4} and \ref{fig:fig5} that the normalized wave phase velocity on given dispersion curve is equal to its label $\mathsf{M_{\sf A}}$).  Therefore, unstable perturbations are frozen in the flow and consequently they are vortices rather than waves.  This observation has a firm physical ground as the KH instability in hydrodynamics deals with unstable vortices.

\section{Discussion and conclusion}
\label{sec:concl}
In this paper, we have studied the condition under which MHD modes traveling on a high-temperature solar surge can become unstable against the Kelvin--Helmholtz instability.  Our model for the surge is a magnetic cylindrical flux tube that might be both untwisted and twisted one.  In the latter case the twist of the surge's magnetic field is characterized by the ratio of azimuthal field component at the inner surface of the tube to its longitudinal component.  The magnetic field of surrounding plasma is considered to be homogeneous (straight magnetic filed lines).  Plasma densities in both the media are also assumed to be homogeneous.  In deriving the dispersion relation of the normal MHD modes propagating on an untwisted flux tube, we have employed two approaches; the first one anticipates that the jet and its environment are compressible plasmas, and the second one, bearing in mind that most of the solar high-temperature surges are observed as hot media, assumes an incompressible jet surrounded by cool/warmer plasma depending on the temperature of the environment.  The dispersion relations of the MHD modes traveling on such configurations are well known and at appropriate plasma and magnetic field characteristics (density contrast $\eta$, magnetic fields' ratio $b$, reduced plasma betas (i.e., ratios $c_{\rm ci}^2/v_{\rm Ai}^2$ and $c_{\rm ce}^2/v_{\rm Ae}^2$, respectively) the derived dispersion curves of the unstable kink mode (mode number $m = 1$) and its instability growth rates turn out to be very similar.  Moreover, critical Alfv\'en Mach numbers for KH instability onset are practically very close: the kink mode propagating on the jet becomes unstable if the jet speed is bigger than ${\approx}1060$~km\,s$^{-1}$ in the case that both media are compressible plasmas, and $1070$~km\,s$^{-1}$ if we treat the jet as incompressible plasma and its environment as a cool medium.  Such speeds, however, are inaccessible for our EUV solar surge. The same conclusion holds for the kink ($m = 1$) mode propagating on a twisted flux tube -- the lowest critical velocity computed at $\varepsilon = 0.025$ is $1072$~km\,s$^{-1}$ -- too high to be detected in any solar surge.  Note that all critical Alfv\'en Mach numbers or equivalently jet speeds critically depend on two parameters, the density contrast, $\eta$, and the background magnetic field, $\bm{B}_{\rm i}$, or more precisely, on the Alfv\'en speed $v_{\rm Ai}$ inside the magnetic flux tube.  Let us consider as a `case study' a high-temperature surge with $n_{\rm i} = 5 \times 10^{10}$~cm$^{-3}$ and $T_{\rm i} = 2.8 \times 10^6$~K \citep{nin}, immersed in a magnetic field of $10$~G and surrounded by plasma having electron number density $n_{\rm e} = 4.8 \times 10^{10}$~cm$^{-3}$ and electron temperature $T_{\rm e} = 3.0 \times 10^6$~K.  Then, we will get a surge--environment configuration with a very low density contrast, $\eta = 0.96$, and Alfv\'en speed $v_{\rm Ai} = 97.5$~km\,s$^{-1}$.  Under these circumstances, the critical speed for the kink mode in an untwisted jet would be equal to $190$~km\,s$^{-1}$.  If we treat both media as incompressible plasmas (which is truly reasonable bearing in mind that their plasma betas are equal to $4.87$ and $5.81$, respectively), then that critical speed increases to $200.4$~km\,s$^{-1}$.  In the case of a twisted flux tube with a weak magnetic field twist parameter, $\varepsilon = 0.025$, the critical speed although a little bit lower, namely equal to $199.8$~km\,s$^{-1}$, is still at the upper limit of accessible surge's speeds.  To be fair, we must exclude the kink ($m = 1$) mode as a MHD wave that can become unstable against the KH instability as propagating along high-temperature solar surges.

The real breakthrough of our study is the finding that the $m = -2$ and $m = -3$ MHD modes can become unstable at sub-Alfv\'enic jet's critical speeds: for our surge under consideration, they are in the ranges of $95.5$--$105$~km\,s$^{-1}$ for the $m = -2$ MHD mode and of $93.5$--$103$~km\,s$^{-1}$ for the $m = -3$ one at magnetic field twist parameter $\varepsilon$ having values between $0.025$ and $0.4$.  As seen from Figs.~\ref{fig:fig4} and \ref{fig:fig5}, the KH instability can occur in four $k_za$-windows, whose location and width is determined by the magnetic field twist parameter, $\varepsilon$.  From an observational point of view, it seems that the unstable $m = -2$ and $m = -3$ MHD modes whose wavelengths lie in the fourth instability window, can really be detected.  The evaluated wave growth rates in the range of several dozen inverse milliseconds look reasonable, but that must be validated (or renounced) by future attempts to image the KH instability in jets like solar surges.  Thus, the answer to the rhetoric question of \cite{rus} paper's title ``Do surges heat the corona?'' is still open -- for a successful modeling of such a phenomenon like the KH instability of MHD modes traveling on solar jets, one needs more reliable observational data for jets' parameters (and their environments) and especially for the background magnetic field.  We do hope that the current observational spacecrafts and future ones like the NASA \emph{Solar Probe Plus\/} and ESA-NASA \emph{Solar Orbiter\/} will provide us invaluable information for our Sun.

%
\acknowledgments
This work was supported by the Bulgarian Science Fund and the Department of Science \& Technology, Government of India Fund under Indo-Bulgarian bilateral project CSTC/INDIA 01/7, /Int/Bulgaria/P-2/12.  We are deeply grateful to the referee for his/her valuable comments and suggestions for improving and clarifying the paper's content.  We are also indebted to Dr.~Snezhana Yordanova for drawing one figure.


%

\end{document}